\documentclass[prl,floatfix,reprint]{revtex4-2}

\usepackage{amsmath}
\usepackage{amssymb}
\usepackage{graphicx}
\usepackage{hyperref}
\usepackage{bm}

\begin{document}

\title{Superbunched radiation of a tunnel junction due to charge quantization}

\author{Steven Kim and Fabian Hassler}
\affiliation{Institute for Quantum Information, RWTH Aachen University, 52056 Aachen, Germany}
\date{June, 2024}
\begin{abstract}
A chaotic light source is characterized by the fact that many independent emitters radiate photons with a random optical phase.
This is similar compared to a tunnel junction where many independent channels are able to emit photons due to a coupling to an electromagnetic environment.
However, in a recent experiment it has been observed that a tunnel junction can deviate from the expectation of chaotic light and is able to emit strongly correlated, superbunched photons.
Motivated by this, we study the correlation of the radiation and show that the superbunching originates from the emission of multiple photons which is possible due to the quantization of charge.
\end{abstract}
\maketitle

\textit{Introduction}.---%
Electrons at low temperature that traverse a constriction biased at a finite voltage $V$ are able to emit photons up to a frequency $eV/\hbar$ due to a coupling to an electromagnetic environment~\cite{lesovik:97}.
Microscopically, this radiation is emitted due to current fluctuations arising from shot-noise caused by the partitioning at the constriction \cite{blanter:00}.
For a quantum point contact, the electron transport is anti-bunched \cite{lesovik:89}. 
It has been shown that for a single transport channel this correlation can be transferred to the emitted photons~\cite{beenakker:04,lebedev:10,fulga:10}.
The correlation of the photons is encoded in the second-order coherence [$a(\tau)$ the annihilation operator of a photon at time $\tau$]
\begin{equation}
g^{(2)}(\tau) = \frac{\langle a^\dagger(0) a^\dagger(\tau) a(\tau) a(0)\rangle}{\langle a^\dagger a\rangle^2},
\end{equation}
describing the correlation between an initially emitted photon and a photon at a later time $\tau$,  where $g^{(2)}(0)<1$ indicates anti-bunching \cite{otten:15}.

However, due to their bosonic nature photons are typically bunched with $g^{(2)}(0) > 1$. 
In fact, a broad class of light sources emit so called `\emph{chaotic light}' that exhibit the special value of $g^{(2)}(0) = 2$ \cite{loudon}. 
Chaotic light is realized in a situation where many independent emitters radiate photons with random optical phases.
At first sight, a tunnel junction, \emph{i.e.}, a constriction with many channels where all the transmission probabilities $D_n \ll1$, is expected to produce chaotic light \cite{beenakker:01, portier:10}.
This is due to the fact that the electron transport in different channels are independent, the transmission of electrons is rare and Poissonian, and the optical phase of the emitted radiation is random.
However, a recent experiment~\cite{leon:19} has shown that a tunnel junction can act as a source of highly correlated light with $g^{(2)}(0)>2$, dubbed `\emph{superbunching}'~\cite{wei:22, ye:22}.
Motivated by this result, we theoretically study the correlation of the radiation emitted by a tunnel junction.

Here, we show that a tunnel junction acts as a chaotic light source only for weak light-matter interaction. 
At stronger interaction, superbunching of photons is predicted by processes where a single electron emits a cascade of multiple photons.
On a fundamental level, the emission of a cascade of photons originates from the quantization of charge making the transport of electrons a point process~\cite{jin:15, esteve:18}.
For a voltage $V > m \hbar \Omega/e$, a single electron is able to emit up to $m$ photons in a single event which is reflected in a large value of $g^{(2)}(0)$.
Note that cascade effects leading to bunching have been studied before in the context of electron transport in molecules \cite{koch:05,belzig:05,gustavsson:09}.
Similar effects have been studied in voltage biased Josephson junctions where a single Cooper pair can emit multiple photons while tunneling \cite{menard:22, lang:21, arndt:22}. 
Note, however, that in the superconducting context the phase of the emitted photons is locked to the phase of the superconducting condensate such that a superconducting junction does not serve as a chaotic light source even at weak light-matter interaction.

The article is organized as follows.
We start by introducing the model and derive a rate equation in the tunnel limit $D_n \ll 1$ that describes the time evolution of the photon occupation in the resonator.
We also take single photon loss due to the coupling to an environment into account.
From this, we determine the stationary state and the second-order coherence.
We show regimes where the second-order coherence exceeds the chaotic value of $g^{(2)}(0)=2$, discuss the impact of temperature, and point out that the strong correlations arise due to the quantization of charge.

\textit{The model}.---%
\begin{figure}[tb]
	\centering
	\includegraphics{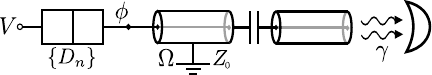}
	\caption{%
	The setup is composed of a voltage biased tunnel junction with transmission probabilities $D_n \ll 1$ in series with a microwave resonator that is capacitively coupled to a transmission line for readout purposes.  The microwave resonator is at frequency $\Omega$ with characteristic impedance $Z_0$. The variable $\phi$ is the node flux describing the voltage $(\hbar/e)\dot \phi$ over the resonator. The transmission line introduces damping; where the photons are lost at rate $\gamma$ into the detector.
}\label{fig:setup}
\end{figure}
We study the setup displayed in Fig.~\ref{fig:setup}.
It consists of a single-mode resonator with the resonance frequency $\Omega$ that can be described by the Hamiltonian $H_\Omega=\hbar \Omega a^\dagger a$.
The voltage across the resonator $(\hbar/e) \dot \phi$ is given by the time-derivative of the node flux $\phi  = \sqrt{\alpha}(a+a^\dagger)$, where $\alpha = (e^2/\hbar)Z_0$ describes the strength of the light-matter coupling with the characteristic impedance $Z_0$ of the resonator. It corresponds to the displacement of the resonator when a single electron tunnels through the junction \cite{ingold:92}.
Due to the capacitive coupling to a readout transmission line, the photons are lost at a rate $\gamma$.

The resonator is in series with a tunnel junction that is biased with a DC voltage $V$. The junction is modeled by a left [L] and right [R] electronic reservoir with the Hamiltonian
\begin{equation}
	H_0 = -i \hbar \sum_n v_n \int \!dx\, \bigl( \psi_{L,n}^\dagger \partial_x \psi_{L,n} + \psi_{R,n}^\dagger \partial_x \psi_{R,n}\bigr) \,;
\end{equation}
here, $v_n$ denotes the velocity of the fermionic mode $\psi_n(x)$ of the $n$-th transport channel \cite{Note1}. The leads are coupled by the tunneling Hamiltonian
\begin{equation}
	H_t = e^{i\phi} \sum_n w_n \psi^\dagger_{R,n}(0)\psi_{L,n}(0) + \text{H.c.}
\end{equation}
where the coupling $w_n$ is connected to the tunneling probability $D_n= |w_n|^2/(2\pi \hbar^2 v_n^2) \ll 1$ \cite{eckern:84}.
Due to the DC voltage bias, the electrons in the leads are distributed as $f_L(\epsilon) = f(\epsilon - eV)$ and $f_R(\epsilon) = f(\epsilon)$ with the Fermi-Dirac distribution $f(\epsilon) = [\exp(\epsilon/k_BT)+1]^{-1}$ at temperature $T$ with the Boltzmann constant $k_B$.

We are interested in obtaining an effective theory for the photon population $P_n$, \emph{i.e.}, the probability that the resonator contains $n$-photons.
As we are in the tunneling limit, we can perform perturbation theory in $H_t$.
The tunneling Hamiltonian couples the displacement operator $e^{i\phi}$ [of the resonator] to the particle-current operator $I= (-i/\hbar)  \sum_{n} w_n \psi^\dag_{R,n}(0)\psi_{L,n}(0) + \text{H.c.}$ [of the electrons forming the bath].
The transition rate $\Gamma^D_{n,n'}$ from the occupation $n'$ to $n= n'+m$ of the resonator can be obtained by Fermi's golden rule. It is given by  $\Gamma^D_{n,n'} = B(m \Omega) | \langle n| e^{i\phi} |n'\rangle|^2$ with $B(\Omega) = \int \!dt \, \langle I(t)I(0)\rangle e^{i\Omega t}$  the [unsymmetrized] shot-noise power, see Ref.~\cite{suppl} for details.
This results in the rate equation
\begin{equation}\label{eq:rate}
	\dot{P}_{n} = \mathcal{L}_D P_n =  \sum_{n' \neq n} \bigl( \Gamma^D_{n,n'}P_{n'}-\Gamma^D_{n',n}P_{n}\bigr)
\end{equation}
with the rate matrix $\mathcal{L}_D$ describing the effect of the tunnel junction on the occupation of the resonator.

The shot-noise power reads $B(\Omega) = D (\Omega+eV/\hbar)n_B(\hbar\Omega + eV) + D(\Omega-eV/\hbar)n_B(\hbar \Omega - eV)$ and depends on the Bose-Einstein distribution $n_B(\epsilon) = [\exp(\epsilon/k_BT)-1]^{-1}$ and the total transmission probability $D = \sum_n D_n$.
The origin is the granularity of the electron charge, as realized by Schottky \cite{schottky:18}, which is the vital ingredient that allows to produce superbunched radiation, see below. 
Note that all transmission channels contribute independently to the dynamics by $D$.
It connects to the DC current through the device via the Landauer formula $I = D(e^2/h)V$, \emph{i.e.}, the conductance is given by $De^2/h$.
Also, the current through the tunnel junction is Poissonian such that no intrinsic electronic correlations can be transferred to the photons.

For $m>0$ [$m<0]$, $B(m\Omega)$ describes emission [absorption] of energy by the conductor~\cite{lesovik:97, aguado:00, marquardt:10}.
At zero temperature, $B(m\Omega)>0$ only if $eV> m\hbar\Omega$.
At finite temperature the rate can be non-zero also for $eV < m\hbar\Omega$.
But, this is suppressed because $B(m\Omega) \propto \exp[-(m\hbar\Omega - eV)/k_BT]$ in this case.
Note that in Refs.~\cite{schull:09,belzig:14}, it has been shown that over-bias emission of photons is also possible at zero temperature.
However, this is due to co-tunneling and thus a higher-order process that scales with $D^2$~\cite{nazarov:06}.

The transition rates are scaled by the Franck-Condon factors [for $m=n -n' > 0$]
\begin{equation}\label{eq:franck}
	\langle n| e^{i\phi} |n'\rangle = \frac{e^{-\alpha/2}(i\sqrt{\alpha})^m}{m!}\sqrt{\frac{n!}{n'!}}\,{}_{1\!}F_1(-n', m+1, \alpha)\,;
\end{equation}  
note that matrix elements for $m<0$ can be obtained by replacing $n\leftrightarrow n'$.
The first factor in \eqref{eq:franck}, that is independent of the state $n$, can be interpreted as a renormalization of $B(m\Omega)$ akin to the Debye-Waller factor. 
The remainder is the transition matrix-element $\langle n|a^\dag{}^m\,{}_{1\!}F_1(-a^\dagger a, m+1, \alpha)|n'\rangle = \sqrt{n!/n'!} \,{}_{1\!}F_1(-n', m+1, \alpha)$ that increases the number of photons by $m$.

To lowest order in the light-matter coupling $\alpha = (e^2/\hbar)Z_0$, we find ${}_{1\!}F_1=1$ such that the Franck-Condon factors  describe the pure creation [absorption] of $m$ photons for $m>0$ [$m<0$] in the resonator in a cascade. 
The hypergeometric function ${}_{1\!}F_1$ becomes relevant at elevated light-matter interaction $\alpha$.
Physically, it arises due to the back-action of the resonator onto the tunnel junction when the voltage across the resonator $(\hbar/e)\dot \phi$ becomes finite and impacts the voltage across the junction.
This back-action can be exploited to achieve a single photon source with $g^{(2)}(0)<1$\cite{gramich:13,rolland:19, esteve:18}.
Note, however, that due to the chirality of quantum hall edge channels it is possible to suppress the back-action such that ${}_{1\!}F_1=1$ for all $\alpha$ \cite{private}.
This is beneficiary for the oberservation of superbunching, see below.

The last ingredient of the modelling is the coupling of the resonator to the detector with rate $\gamma$ and at temperature $T$.
It leads to the absorption and emission of photons given by the rate equation $\dot{P_n} = \mathcal{L}_\gamma P_n$ with the rates
\begin{equation}\label{eq:L_Z}
	\Gamma^\gamma_{n,n'} = \gamma n_0 n\delta_{n,n'+1} + \gamma (n_0+1)(n+1)  \delta_{n,n'-1}
\end{equation}
where $n_0=n_B(\hbar\Omega)$, see \emph{e.g.}~\cite{breuer}. 
The total time-evolution of the probabilities $P_n$ is thus given by $\mathcal{L} = \mathcal{L}_D  +\mathcal{L}_\gamma$. 
It incorporates both the interaction of the resonator with the tunnel junction and the detector.
Note that the phase of the emitted photons of the resonator is arbitrary.
This is due to the fact that the rate matrix $\mathcal{L}$ only acts on the diagonal of the density matrix $\rho = \sum_n P_n |n\rangle\langle n|$.
In this sense, the system consists of many independent sources given by the different channels that emit radiation with a random optical phase. 
Still, we will observe superbunching of the radiation due to the fact that the electronic charge is granular.

\textit{Second-order coherence}.---%
The second-order coherence quantifies the correlation of photons. 
For chaotic light, it can be shown that $g^{(2)}(0) =2$ \cite{loudon}. 
Such light sources include blackbody radiation and emission from an Ohmic resistor, see below. 
We determine the value of the second-order coherence for our system by solving for the stationary probability distribution $P_{s,n}$, fulfilling $\dot P_{s,n} = \mathcal{L}P_{s,n} =0$.
The second-order coherence can then be obtained by $g^{(2)}(0)  = \langle a^\dagger{}^2 a^2\rangle_s/ \langle a^\dagger a\rangle^2_s =  \langle n(n-1)\rangle_s/ \langle n\rangle^2_s  $ with $\langle O\rangle_s = \sum_n P_{s,n} \langle n|O|n\rangle$ \cite{Note2}.

\begin{figure*}
\centering
\includegraphics[]{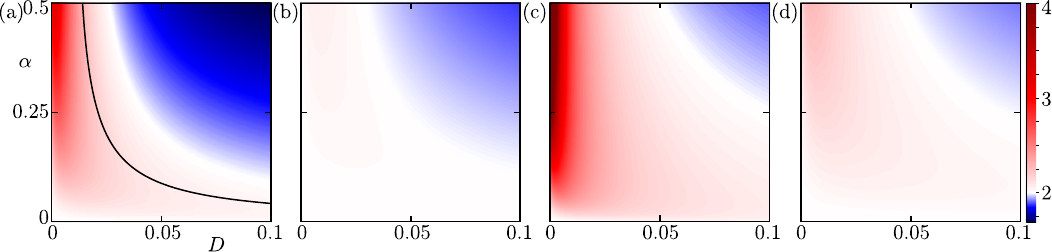}
\caption{
	Second-order coherence $g^{(2)}(0)$ as a function of the light-matter coupling $\alpha$ and conductance $D$ for $eV=2.2\hbar\Omega$ [up to 2 photon process possible)] in (a) and (b) and  for $eV=4.2\hbar\Omega$ [up to 4 photon process possible] in (c) and (d); with $\gamma=0.1\Omega$ and $n_0=0.01$ [$n_0=0.1$] in (a),(c) [in (b),(d)].  The superbunched region is indicated in red.
	Note that in (c) values up to $5.8$ are present [the colorbar is maxed out at 4 for better visibility]. The black line in (a) indicates the transition line $g^{(2)}(0)=2$ at zero temperature; at finite temperature, the region of superbunching is larger. In general for larger voltages, both the region of superbunching as well as the values of $g^{(2)}(0)$ are increased.
}\label{fig:g2}
\end{figure*}

Results for the second-order coherence obtained by a numerical simulation are shown in Fig.~\ref{fig:g2} for different parameters. 
It can be seen that especially at low-temperature, $D$ small, and large voltages superbunched radiation with $g^{(2)}(0)>2$ is produced. 
The smallness of the transmission $D$ is required such that the superbunching events are rare and separate in time.
Additionally, the light-matter interaction has to be sufficiently large such that multi-photon processes are present at all \cite{menard:22}.

We would like to obtain further analytical insights into the superbunching effect. 
For the following, we concentrate on zero temperature [$n_0 = 0$] and choose a voltage with $2\hbar \Omega < eV < 3\hbar \Omega$ such that only one- and two-photon cascade events are present, corresponding to $m=1,2$.
The shot-noise power for these processes is given by $B(m\Omega) = eV/\hbar-m\Omega$.

First, we focus on the one-photon dynamics.
In this situation, single photons are lost with a rate $\gamma$ while they are predominantely generated by the $m=1$ process with the rate $\gamma_g= \alpha e^{-\alpha} D (eV/ \hbar - \Omega) \simeq \alpha e^{-\alpha} D \Omega$, incorporating the Debye-Waller factor.
For convenience, we assume that $\alpha e^{-\alpha} D \ll \gamma/\Omega$ and $\alpha$ small which is fulfilled in the relevant part of Fig.~\ref{fig:g2}. 
In this regime, we have that $\gamma_g \ll \gamma$.
We obtain the approximate rate equation $\dot P_n = \gamma_g [n P_{n-1} - (n+1)P_n] + \gamma[(n+1) P_{n+1}- n P_n]$.
It yields the Gaussian contribution since it only includes single photon transitions.
The stationary state is an effective thermal state with the mean photon number in the resonator $\bar n = \langle a^\dagger a \rangle_s = 1/(\gamma/\gamma_g-1) \approx \gamma_g/\gamma \ll 1$. 
Because of this, we have $g^{(2)}(0)=2$ to this order. 
Note that the same rate equation is also obtained for the situation of an Ohmic resistor when the discreteness of charge can be neglected such that  $e^{i\phi} \mapsto 1 +i\phi$.
This shows that the superbunched radiation is a property of the quantization of charge such that a single electron emits a cascade of photons. 

To obtain the cascade effect, we have to calculate the next-order correction in $\alpha e^{-\alpha} D \ll \gamma/\Omega$ which corresponds to $\bar n \ll 1$. 
There are two new contributions in this order. 
The first arises due to the expansion of the hypergeometric function ${}_{1\!}F_1(-n',2,\alpha)= 1 - \frac12 \alpha n'$ for $m=1$, that accounts for the back-action.
The second process that is important to this order is the two photon cascade described by the $m=2$ process. Note that  for this process, the back-action can be neglected, as it is subdominant. 
As shown in \cite{suppl}, the average photon number in this approximation is given by
\begin{equation}\label{eq:adaggera}
\langle a^\dag a \rangle_s = \bar n -2 \alpha \bar n^2 + \kappa \bar n( 1+ 2 \bar n) + \mathcal{O}(\bar n^3) 
\end{equation}
with the unperturbed photon number $\bar n =  1/(\gamma/\gamma_g-1)$ and $\kappa = \alpha (eV-2\hbar\Omega)/(eV-\hbar\Omega)$ the ratio of the two-photon to the one-photon rate.
Note that the two-photon process [last term proportional to $\kappa$] increases the photon number while the back-action [second term proportional to $\alpha$] counteracts this.

In order to obtain the second-order coherence at zero time-delay, we additionally need the photon-photon correlation
\begin{equation}\label{eq:corr}
	\langle a^\dagger{}^2a^2 \rangle_s = 2 \bar{n}^2 - 2\alpha \bar n^2 + \kappa \bar n ( \tfrac12 + 6\bar n) + \mathcal{O}(\bar n^3).
\end{equation}
As above, the first term is the Gaussian contribution due to the unperturbed solution. 
The second term incorporates the effect of the back-action while the last term is due to the two-photon cascade process.
At small $\bar n$, the term $\tfrac12\kappa\bar n$ dominates that arises due to the emission of two photons by a single electron. This term leads to superbunching which will be explained in the following.

The second-order coherence is given by the ratio $g^{(2)}(0) = \langle a^\dagger{}^2a^2\rangle_s/\langle a^\dagger a\rangle_s^2$. In order to obtain compact expressions, we concentrate on voltages close to the two-photon threshold $eV \gtrsim 2 \hbar \Omega$ such that $\kappa \ll 1$.
In the limit $\kappa, \bar n \ll 1$, we obtain the second-order coherence
\begin{equation}\label{eq:g2}
	g^{(2)}(0) \approx 2 -2 \alpha  + \frac{\kappa}{2 \bar n}= 2  - 2 \alpha + \frac{e^\alpha \hbar\gamma(eV-2\hbar\Omega)}{2D(eV-\hbar\Omega)^2},
\end{equation}
see~\cite{suppl} for a more general result.
The first term [$g^{(2)}(0) = 2$] describes the Gaussian contribution and encodes the intuitive expectation that a tunnel junction acts as a chaotic light source.
The second term, being proportional to $\alpha$, stems from the back-action expressed by the hypergeometric function ${}_{1\!}F_1(-n',2,\alpha)$ of the $m=1$ process.
Note that it always lowers the second-order coherence and counteracts the observation of superbunching.
If the back-action is suppressed, \emph{e.g.}, by employing the chirality of quantum Hall edge channels, this term is absent. 

The last term leads to superbunching. It arises as the ratio of $\frac12 \kappa \bar n$, due to the cascaded emission of two-photons produced by a single electron, to $\bar n^2$ due to the unperturbed photon number. Being inversely proportional to $\bar n$,
it yields a large positive contribution and makes the observation of superbunching possible \cite{Note3}.
The cascade events scale with $g^{(2)}(0) \propto D^{-1}$ such that the superbunching effect is strongest when the transmission probability is small.
Then, the superbunched photons are created separate in time and can be emitted by the resonator before a new bunch is created.
This effect remains observable also at voltages where higher order cascades are possible.

The superbunching persists to finite temperatures.
In this case, the divergence in the limit $D \rightarrow 0$  is cured and the systems occupies a thermal state with the average occupation $n_0$ and $g^{(2)}(0)=2$~\cite{flindt:19}.
At elevated conductance $D$, the behavior \eqref{eq:g2} remains unchanged.
For small temperature, the superbunching is maximal at the crossover scale with $n_0 = \bar n \approx \alpha e^{-\alpha} D(eV-\hbar \Omega)/\hbar\gamma$.
As a result, for the two-photon cascade with $2\hbar \Omega < eV < 3 \hbar \Omega$ a maximal value of $g^{(2)}(0)\approx \alpha (eV - 2 \hbar \Omega)/[2 n_0 (eV-\hbar \Omega)] \simeq \alpha/n_0 $ can be found at $D^* \approx \hbar\gamma n_0 /[\alpha e^{-\alpha} (eV-\hbar\Omega)]$.

The results obtained in this work can be measured in setups, Fig.~\ref{fig:setup},  operating at microwave frequencies that are already available.
Today's experiments can achieve resonators with a characteristic impedance $Z_0 \approx  1\,\mathrm{k}\Omega$, frequency $\Omega \approx 33$\,GHz, operated at temperatures $T \approx 50$\,mK~\cite{private}.
This yields $\alpha \approx 0.24$ for the light-matter coupling and an Bose-Einstein occupation of $n_0\approx 5\times 10^{-3}$. For these parameters at the optimal value $D^*$, we obtain $g^{(2)}(0) \approx 8$ for the two-photon cascade.
Note that our findings provide insights into the intriguing experiment of Ref.~\cite{leon:19} which measured  $g^{(2)}(0)\approx 70$  at optical frequencies.
For optical wavelengths, the zero temperature result~(\ref{eq:g2}) is applicable which explains the large values of the second-order coherence at small $D$ with $g^{2}(0)\propto D^{-1}$. 
Due to the absence of a cavity in Ref.~\cite{leon:19}, for an estimate, we set $\gamma \simeq \Omega$ and obtain $g^{(2)}(0) \simeq 100$ at the experimental value $D \simeq 10^{-3}$.

\textit{Conclusion}.---%
To conclude, we have studied the correlation of the emitted radiation by a tunnel junction.
Intuitively, it is expected that the tunnel junction acts as a chaotic light source because many independent channels emit photons with a random optical phase.
However, we have shown that the quantization of charge yields non-Gaussian dynamics and makes the emission of multiple photons in a cascade during a single transmission event possible.
Because of this, the photons are highly correlated and superbunching with $g^{(2)}(0) > 2$ can be observed \cite{leon:19}. 
Additionally, we have analyzed the effect of temperature on the correlation, studied the emission of photon pairs analytically, and provided numerical results for higher order processes.
The correlations peaks at a crossover between the thermal occupation of the environment and the effective thermal state due to the tunnel junction at $D^*=  \hbar\gamma n_0 /[\alpha e^{-\alpha} (eV-\hbar\Omega)]$.
While emitters with $g^{(2)}(0)>1$ are typically considered as classical light sources, we have shown that the strong correlations arise due to quantum effects, in particular the quantization of charge.

We acknowledge valuable discussions with C. Altimiras and O. Ghazouani Gharbi.
This work was supported by the Deutsche Forschungsgemeinschaft (DFG) under Grant No.\ HA~7084/8--1.

\onecolumngrid\clearpage\appendix
\setcounter{equation}{0}\renewcommand\theequation{S\arabic{equation}}
\setcounter{figure}{0}\renewcommand\thefigure{S\arabic{figure}}\renewcommand\theHfigure\thefigure
\setcounter{table}{0}\renewcommand\thetable{S\Roman{table}}\renewcommand\theHtable\thetable
\setcounter{page}{1}
\makeatletter\let@environment{thebibliography}{NAT@thebibliography}\makeatother

\section*{Supplemental Material}
\begin{quote}
Here, we show detailed calculations on how to obtain the results of this paper.
In the first part of this supplement, we go beyond Fermi's golden rule and demonstrate how to use the Bloch-Redfield equation to obtain a Lindblad master equation in the rotating frame that is equivalent to rate equation Eq.~\eqref{eq:rate}.
Then, we calculate the second-order coherence for zero time delay $g^{(2)}(0)$ by determining the stationary state in the second part of the supplement.
We also show that $g^{(2)}(\tau)$ decays monotonically to $1$ on the timescale $\gamma^{-1}$ which is why $g^{(2)}(0)$ is sufficient to see whether superbunching is present or not.
\end{quote}
\subsection{Rate equation}
Our total system consists of a resonator with Hamiltonian $H_\Omega = \hbar \Omega a^\dagger a$ that interacts with the electronic bath of the leads $H_0 = -i \hbar \sum_n v_n \int \!dx\, \bigl( \psi_{L,n}^\dagger \partial_x \psi_{L,n} + \psi_{R,n}^\dagger \partial_x \psi_{R,n}\bigr)$ via the tunneling Hamiltonian $H_t = e^{i\phi}  \sum_n w_n \psi^\dagger_{R,n}(0)\psi_{L,n}(0) + \text{H.c.}$ with $\phi = \sqrt\alpha(a+a^\dagger)$.
Our goal is to find an effective description of the density matrix of the resonator $\rho$ by tracing out the electrons $\rho_e$.
Because the tunnel probabilities  $D_n = |w_n|^2 / (2 \pi \hbar^2 v_n^2 ) \ll 1$ are small for a tunnel junction, the effect of $H_t$ can be incorporated by perturbation theory.

We go over to the interaction-representation with respect to the Hamiltonian $H_0 + H_\Omega$. 
The time evolution of $\rho$ is described by the Bloch-Redfield equation
\begin{equation}\label{eq:redfield}
\dot \rho = -\frac{1}{\hbar^2}\int_0^\infty d\tau \, \text{tr}_e[H_t(t),[H_t(t-\tau),\rho(t)\otimes \rho_e]],
\end{equation}
were $\text{tr}_e$ is the trace over the electronic degrees of freedom.
Here, we have assumed that the density matrix of the total system can be decomposed as $\rho(t)\otimes \rho_e$ which is valid in the tunnel limit.
Also, we have assumed that correlations in the electronic bath decay much faster than the timescale over which the system evolves.
This is commonly known as the Born-Markov approximation.

In the interaction picture, the bosonic operators transform according to $a(t) = a e^{-i\Omega t}$. To perform a rotating wave approximation [RWA], it is useful to employ the identity 
\begin{equation}
	e^{i\phi(t)}= e^{i\sqrt\alpha[a(t)+a^\dag(t)]}= \sum_{m\in\mathbb{Z}} i^m A_m e^{im\Omega t}, \qquad A_m = \begin{cases}
		\frac{e^{-\alpha/2} \alpha^{m/2}}{m!} a^\dag{}^m \,{}_{1\!}F_1(-a^\dag a, m+1, \alpha), & m\geq 0 \\
	A_{-m}^\dag,& m <0. \end{cases}
\end{equation} 
Note that the tunneling Hamiltonian can be written as $H_t(t) = C(t)\sum_m i^m A_m  e^{im\Omega t}$ where $C(t) = \sum_n \bigl[w_n \psi^\dagger_{R,n}(0;t)\psi_{L,n}(0;t) + \text{H.c.}\bigr]$.
The Bloch-Redfield equation becomes
\begin{equation}
\dot \rho(t) = -\frac1{\hbar^2} \int_0^\infty \!\! d\tau \sum_{m,m'}\Bigl[ G(\tau)e^{-i m'\Omega \tau} A_m  A_{m'} e^{i(m+m')\Omega t}\rho(t) - G(-\tau) e^{-i m'\Omega \tau} A_m \rho(t) A_{m'} e^{i(m+m')\Omega t} + \text{H.c.}\Bigr]
\end{equation}
with the correlator $G(\tau) = \text{tr}_e [C(\tau)C(0)\rho_e] = \langle C(\tau)C(0)\rangle = \hbar^2\langle I(t)I(0)\rangle$, where $I = (-i/\hbar)\sum_n w_n \psi^\dag_{R,n}\psi_{L,n} +\text{H.c.}$ is the particle current operator as given in the main text.
Here, we are able to perform the RWA by neglecting fast oscillating terms with $m'\neq -m$.
The time evolution of the density matrix is then given by 
\begin{equation}
\dot \rho(t) = - \sum_{m} \int_0^\infty \!\! d\tau
\Bigl[ G(\tau)e^{i m\Omega \tau} A_m  A_m^\dag \rho(t) - G(-\tau) e^{i m\Omega \tau} A_m \rho(t) A_{m}^\dag  + \text{H.c.} \Bigr].
\end{equation}

We are left with evaluating the one sided Fourier transforms of the correlator $B(m\Omega) = 2 \int_0^\infty G(\tau)e^{im\Omega \tau} d\tau$  with
\begin{equation}
G(\tau) = \frac{1}{\hbar^2}\sum_{n} |w_n|^2 \Biggl[\int\!\!\int\!\! \frac{dk\,dk'}{(2\pi)^2} \langle c_{R,k,n}^\dag c_{R,k,n} c_{L,k',n} c^\dag_{L,k',n} \rangle e^{i(\omega_{k,n}-\omega_{k',n})\tau}+ \text{H.c.}\Biggr];
\end{equation}
here, we have introduced $\psi(x) = \int \tfrac{dk}{2\pi} e^{ikx} c_{k}$ and $\omega_{k,n} = v_n k$ the frequency set by the electronic bath $H_0$. 
Generally, the Fourier transform yields a real and imaginary part.
The imaginary part leads to an energy shift of the Hamiltonian, also called Lamb-shift, which will be neglected.
The real part can be evaluated to 
\begin{equation}
B(m\Omega) = 2 \pi\sum_n  |w_n|^2 N_n^2 \int\!\!\int\!\!d\omega_kd\omega_k'\Bigl\{ f_R(\hbar\omega_k)[1-f_L(\hbar\omega_k')]\delta(m\Omega + \omega_k-\omega_k') + [1-f_R(\hbar\omega_k)]f_L(\hbar\omega_k')\delta(m\Omega - \omega_k+\omega_k')  \Bigr\}
\end{equation}
with $N_n = 1/(2\pi\hbar v_n)$ the density of states, $f_R(\epsilon) = f(\epsilon)$, $f_L(\epsilon) = f(\epsilon-eV)$, and $f(\epsilon) = [\exp(\beta\epsilon)+1]^{-1}$ the Fermi-Dirac distribution at the inverse temperature $\beta =1/k_BT$ and $k_B$ the Boltzmann constant. Note that $B(m \Omega)$ is the [unsymmetrized] shot-noise power \cite{blanter:00}.

The integral yields $B(\Omega) = D(\Omega-eV/\hbar)n_B(\hbar\Omega-eV) +  D(\Omega+eV/\hbar)n_B(\hbar\Omega+eV)$ with $D=\sum_n D_n = \sum_n |w_n|^2/(2\pi\hbar^2v_n^2)$ and the Bose-Einstein distribution $n_B(\epsilon) = [\exp(\beta\epsilon)-1]^{-1}$.
In total the Bloch-Redfield equation becomes a Lindblad master equation with
\begin{equation}
\dot{\rho} = \sum_{m=-\infty}^\infty B(m\Omega) \bigr[A_m\rho A_m^\dag -\tfrac12(A_m^\dag A_m \rho + \rho A_m^\dag A_m )\bigr].
\end{equation}
Introducing $A_m = e^{-\frac{\alpha}{2}}(i\sqrt\alpha)^m b_m^\dagger/m!$ where $b_m = {}_{1\!}F_1(-a^\dag a, m+1, \alpha)a^m$, the Lindblad equation can be written as $\dot \rho = \mathcal{L}\rho$ with the Liouvillian 
\begin{equation}
	\mathcal{L} =  e^{-\alpha}  \sum_{m=1}^\infty \frac{\alpha^m}{(m!)^2} \Bigl\{ B(m\Omega) \mathcal{J}[b_m^\dagger] + B(-m\Omega)\mathcal{J}[b_m] \Bigr\}.
\end{equation}
We have defined the dissipator $\mathcal{J}[L]\rho = L\rho L^\dagger -\tfrac12(L^\dagger L \rho + \rho L^\dagger L)$.
In lowest order of $\alpha$, the jump operator is given by $b_m=a^m$.
Thus, the jump operators describe creation [annihilation] of $m$ photons by a cascade expressed through $b_m^\dag$ [$b_m$] with the rate $e^{-\alpha}\alpha^m B(\pm m\Omega)/(m!)^2$.

The Bloch-Redfield equation, or also the resulting Lindblad equation in the rotating frame, yield the time evolution of the full density matrix.
Thus, they contain more information about the system than the rate equation that is given in the main text.
However, when we only consider the main diagonal of the density matrix $\rho = \sum_n P_n|n\rangle\langle n|$ we immediatly obtain the rate equation given in the main text that can be obtained by Fermi's golden rule \cite{breuer}.

\subsection{Second-order coherence}
Here, we want to derive the results \eqref{eq:adaggera}, \eqref{eq:corr}, and \eqref{eq:g2} from the main text.
We focus on the zero temperature limit and include the relevant terms for the two-photon cascades as described in the main text.
We write $\mathcal{L}\approx\mathcal{L}_1 + \mathcal{L}_2$, see below.
The corresponding rate equations are given by
\begin{equation}
\mathcal{L}_1 P_n = \gamma_g [n P_{n-1} - (n+1)P_n] +  \gamma[ (n+1) P_{n+1}-nP_n]
\end{equation}
and
\begin{equation}
\frac{\mathcal{L}_2 P_n }{\alpha^2e^{-\alpha}}= \frac{B(2\Omega)}{4}[n(n-1) P_{n-2} - (n+1)(n+2)P_n] - B(\Omega)[n(n-1)P_{n-1} - n(n+1)P_n]  ,
\end{equation}
where $\gamma_g = \alpha e^{-\alpha}B(\Omega)$ and $ B(\Omega)=D(eV/\hbar-\Omega)$ at zero temperature.
To solve the rate equation, we treat $\mathcal{L}_2$ as a perturbation.
Then, the probability distribution is given by $P_n \approx P_n^{(0)} + P_n^{(1)}$ where $P_n^{(0)}$ and $P_n^{(1)}$ are solutions of the difference equations $\mathcal{L}_1P_n^{(0)}=0$ and $\mathcal{L}_1P_n^{(1)}=-\mathcal{L}_2P_n^{(0)}$.
Fortunately, both can be solved.
The first rate equation $\mathcal{L}_1P_n^{(0)}=0$ can be solved by $P_n^{(0)} = (1-z)z^n$ with $z=\gamma_g/\gamma \ll 1$ and $\sum_n P_n^{(0)}=1$.
We can write $\mathcal{L}_2P_n^{(0)} = \gamma_g (1-z)z^n F_n$ with
\begin{equation}
F_n =\frac{\alpha e^{-\alpha}}{\gamma_g} \biggl[ \frac{B(2\Omega)}{4 z^2}n(n-1) - \frac{B(2\Omega)}{4}(n+1)(n+2) + B(\Omega) n(n+1)-\frac {B(\Omega)}{z}n(n-1)\biggr].
\end{equation}
The solution of $\mathcal{L}_1P_n^{(1)}=-\gamma_g (1-z) z^n F_n$ is  given by
\begin{equation}
P_n^{(1)} = (1-z)z^n \sum_{m=0}^n \Bigl[ F_m \sum_{k=0}^m\frac{z^k}{n+1-k}\Bigr]  + (1-z)\Bigl[\sum_{m=n+1}^\infty F_m z^m\Bigr]\Bigl[\sum_{m=1}^n \frac{z^m}{n+1-m}\Bigr].
\end{equation}
The full solution reads $P_n = (1+\mathcal{N})P_n^{(0)} +P_n^{(1)}$ where $\mathcal{N}= -\sum_n P_n^{(1)}$ such that $\sum_n P_n = \sum_n P_n^{(0)} =1$.
The Gaussian contribution $P_n^{(0)}$ leads to the average photon number $\bar n = \sum_n nP_n = 1/(\gamma/\gamma_g-1)$.

We want to obtain the second-order coherence in leading order of $\bar n \ll 1$  equivalent to $\gamma_g\ll \gamma$ or also $\alpha e^{-\alpha}D\ll \gamma/\Omega$ as given in the main text.
Thus, we need to evaluate $\langle a^\dagger a \rangle_s = \sum_n nP_n$ and $\langle a^\dagger{}^2 a^2 \rangle_s = \sum_n n(n-1)P_n$ to second order.
We obtain
\begin{equation}
\langle a^\dag a \rangle_s = \bar n -2 \alpha \bar n^2 + \kappa \bar n( 1+ 2 \bar n) + \mathcal{O}(\bar n^3) 
\end{equation}
and
\begin{equation}
	\langle a^\dagger{}^2a^2 \rangle_s = 2 \bar{n}^2 - 2\alpha \bar n^2 + \kappa \bar n ( \tfrac12 + 6\bar n) + \mathcal{O}(\bar n^3).
\end{equation}
with $\kappa = \alpha B(2\Omega)/B(\Omega) = \alpha(eV-2\hbar\Omega)/(eV-\hbar\Omega)$.
This leads to the second-order coherence
\begin{equation}
	g^{(2)}(0) = 2 - 2\frac{\alpha-\kappa+\kappa^2+\kappa^3}{(1+\kappa)^3} +  \frac{\kappa}{2(1+\kappa)^2 \bar n} + \mathcal{O}(\bar n)
\end{equation}
Close to the two-photon threshold $eV\gtrsim 2\hbar\Omega$ we have that $\kappa \ll 1$ which leads to the result stated in the main text.
Here we also see that the two-photon process also yields a constant negative shift which counteracts the superbunching.
However, these are of higher orders in $\kappa$ and renormalize the effect of the back-action as explained in the maint text.

\begin{figure}[tb]
	\centering
	\includegraphics{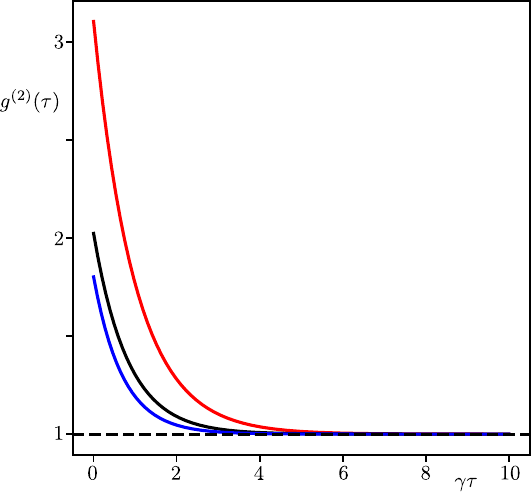}
	\caption{%
	Time dependence of the second-order coherence $g^{(2)}(\tau)$ for $eV=2.2\hbar\Omega$, $\gamma=0.1\Omega$, $\alpha=0.5$, $n_0=0.01$,  and $D=0.001,0.025, 0.5$ in red, black, blue.
	The second-order coherence decays monotonically to $1$ from a starting value $g^{(2)}(0)$ which indicates whether superbunching is present or not.
}\label{fig:g2tau}
\end{figure}
Furthermore, we show results or the time dependence of the second-order coherence $g^{(2)}(\tau)$ in Fig.\ref{fig:g2tau}. We see that $g(\tau)$ is a smooth function with a maximum $g^{(2)}(0)$ at $\tau =0$ that decays on the scale $\gamma^{-1}$ to 1. Because of this, 
$g^{(2)}(0)$ encodes the [all the] crucial information about the correlator; \emph{e.g.}, the Fano factor can be estimated as $F= \langle\!\langle N^2 \rangle\!\rangle/  \langle\!\langle N \rangle\!\rangle = 1+ \gamma\bar n\int\!d\tau [g^{(2)}(\tau)-1] \approx 1 +2\bar n g^{(2)}(0) $\,.

\end{document}